\documentclass[copyright,creativecommons]{eptcs}
 \providecommand{\event}{Developments in Computational Models 2012} % Name of the event you are submitting to
\usepackage{amsmath}
\usepackage{amssymb}
\usepackage{amsthm}

\title{Non-deterministic computation and the Jayne-Rogers Theorem}
\author{
Arno Pauly
\institute{Computer Laboratory\\ University of Cambridge, United Kingdom
\email{Arno.Pauly@cl.cam.ac.uk}}
\and
Matthew de Brecht
\institute{National Institute of Information and Communications
Technology\\Kyoto, Japan\email{matthew@nict.go.jp}}
}
\def\titlerunning{Effective $\Delta_2^0$-measurability}
\def\authorrunning{A.\ Pauly and M.\ de Brecht}

\begin{document} \theoremstyle{definition}
\newtheorem{theorem}{Theorem}
\newtheorem{definition}[theorem]{Definition}
\newtheorem{problem}[theorem]{Problem}
\newtheorem{assumption}[theorem]{Assumption}
\newtheorem{corollary}[theorem]{Corollary}
\newtheorem{proposition}[theorem]{Proposition}
\newtheorem{lemma}[theorem]{Lemma}
\newtheorem{observation}[theorem]{Observation}
\newtheorem{fact}[theorem]{Fact}
\newtheorem{question}[theorem]{Question}
\newtheorem{example}[theorem]{Example}
\newcommand{\dom}{\operatorname{dom}}
\newcommand{\range}{\operatorname{range}}
\newcommand{\cf}{\operatorname{cf}}
\newcommand{\Dom}{\operatorname{Dom}}
\newcommand{\codom}{\operatorname{CDom}}
\newcommand{\id}{\textnormal{id}}
\newcommand{\Cantor}{\{0, 1\}^\mathbb{N}}
\newcommand{\Baire}{\mathbb{N}^\mathbb{N}}
\newcommand{\lev}{\textnormal{Lev}}
\newcommand{\hide}[1]{}
\newcommand{\mto}{\rightrightarrows}
\newcommand{\pls}{\textsc{PLS}}
\newcommand{\ppad}{\textsc{PPAD}}
\newcommand{\finalobj}{\mathbb{F}}
\newcommand{\catprob}{Prob}
\newcommand{\lpo}{\textrm{LPO}}
\newcommand{\uint}{{[0, 1]}}
\newcommand{\Sierp}{Sierpi\'nski }
\newcommand{\name}[1]{#1}

\maketitle

\begin{abstract}
We provide a simple proof of a computable analogue to the Jayne-Rogers Theorem from descriptive set theory. The difficulty of the proof is delegated to a simulation result pertaining to non-de\-ter\-mi\-nistic type-2 machines. Thus, we demonstrate that developments in computational models can have applications in fields thought to be far removed from it.
\end{abstract}

\section{Introduction}
Non-deterministic type-2 machines (NDTMs) were suggested by \name{Ziegler} \cite{ziegler2,ziegler3,ziegler7} as a model for hypercomputation in computable analysis. As demonstrated by \name{Brattka}, \name{de Brecht} and \name{Pauly} \cite{paulybrattka,paulybrattka2}, the strength of various kinds of type-2 non-de\-ter-mi\-nism neatly classifies various important classes of non-computable functions; and a characterization of such classes as those functions computable by certain NDTMs opens up new, simple ways to prove closure properties for them.

A NDTM with advice space $Z$ is a Turing machine with an input tape, an oracle tape, some work tapes and a write-once output tape. The input is an infinite sequence written on the input tape, the oracle tape is initialized with a \emph{guess}, an infinite sequence from the set $Z$. The machine either halts eventually, which is seen as a rejection of the guess, or continues to write an infinite sequence on the output tape. For any valid input there must be an acceptable guess.

Thus, a NDTM naturally computes a multivalued function $f : \subseteq \Cantor \mto \Cantor$. The notion of non-de\-ter-mi\-nistic
 computability is then lifted to arbitrary represented spaces: Some $f : \mathbf{X} \mto \mathbf{Y}$ is non-deterministally computable with advice space $Z$, iff there is an NDTM such that any $p \in \Cantor$ denoting an element of $\mathbf{X}$ is accepted, and every successful computation produces a name for some $y \in f(x)$.

The power of NDTMs severely depends on the advice space. The spaces $\Cantor$ and $\mathbb{N}$ yield incomparable computational power, $\mathbb{N} \times \Cantor$ is more powerful than both, and $\Baire$ again significantly more powerful than $\mathbb{N} \times \Cantor$. The crucial property for us is that the additional computational power of $\mathbb{N} \times \Cantor$ over $\mathbb{N}$ only applies to multivalued functions---any single-valued $f : \mathbf{X} \to \mathbf{Y}$ (with $\mathbf{Y}$ computably admissible) non-de\-ter-mi\-nistically computable with advice space $\mathbb{N} \times \Cantor$ already is non-de\-ter-mi\-nistically computable with advice space $\mathbb{N}$.

We will apply the theory of non-de\-ter-mi\-nistic computations to descriptive set theory. A subset of a metric space is called $\Delta_2^0$, if it is both the countable union of closed sets and the countable intersection of open sets. A function is called $\Delta_2^0$-measurable, iff the preimage of any open set is a $\Delta_2^0$-set. A function will be called $\mathcal{A}$-piecewise continuous, iff there is a countable cover of its domain by closed sets, such that the restriction to any such closed set is continuous.

\begin{theorem}[\name{Jayne} \& \name{Rogers} \cite{jaynerogers}]
\label{theo:jaynerogers}
Let $\mathbf{X}$, $\mathbf{Y}$ be metric spaces. If $\mathbf{X}$ is absolute Souslin-$\mathfrak{F}$, then $f : \mathbf{X} \to \mathbf{Y}$ is $\Delta_2^0$-measurable if and only if it is $\mathcal{A}$-piecewise continuous.
\end{theorem}

For a definition of absolute Souslin-$\mathfrak{F}$, see \cite[Definition 25.4]{kechris}. This criterion is not needed in the computable case, hence we do not give details here.

After the original proof by \name{Jayne} and \name{Rogers} \cite{jaynerogers}, simplifications were provided first by \name{Solecki} \cite{solecki} and then by \name{Motto Ros} and \name{Semmes} \cite{ros}\footnote{See also \cite{ros2}.}. While showing that piecewise continuous functions are $\Delta_2^0$-measurable is straight-forward, the known proofs for the other direction are all somewhat complicated, and non-constructive: The assumption that a function was both $\Delta_2^0$-measurable and not piecewise continuous is taken to a contradiction; there is no construction given for the countable closed cover witnessing the piecewise continuity of a $\Delta_2^0$-measurable function.

Working in the framework of computable analysis, we will provide a computable version of Theorem \ref{theo:jaynerogers}, that is we show how to compute information identifying a function as piecewise continuous from information identifying it as $\Delta_2^0$-measurable. This should not be confused with giving an analogue to Theorem \ref{theo:jaynerogers} in \emph{effective} descriptive set theory, which is investigated in Section \ref{section:Markov}.

\section{Non-deterministic type-2 machines}
It is well-established (\name{Weihrauch} \cite{weihrauchd}) that a good model for computation on objects from analysis is given by Type-2 machines. These essentially are the usual (deterministic) Turing machines with new semantics: The computation runs forever, hence every cell of the infinite tapes is actually accessible. A designated output tape allows the head only to move to the right, so any symbol to the left of the current head position remains unchanged. A well-behaved computation will write on the output infinitely often, and thus produces an infinite output sequence. Type-2 machines should not be confused with the Infinite Time Turing Machines proposed by \name{Hamkins} and \name{Lewis} \cite{hamkins}---both the definitions and their purpose differ significantly.

Just as Type-2 machines are derived from deterministic Turing machines, we can derive non-de\-ter-mi\-nistic Type-2 machines from non-de\-ter-mi\-nistic Turing machines. We assume that the non-de\-ter-mi\-nism is localized, i.e., that all non-de\-ter-mi\-nistic bits are guessed at once. As a Type-2 machine has unlimited time available for the verification, it can utilize an infinite sequence of non-de\-ter-mi\-nistic bits. In addition, we provide our machines with the promise that the guess-sequence is in some fixed set $Z \subseteq \Cantor$---a property not relevant in the classical case.

A crucial difference between Type-1 and Type-2 non-de\-ter-mi\-nism is that the latter increases the computational power, whereas non-de\-ter-mi\-nistic Turing machines can be simulated by deterministic ones. A typical example of a non-computable problem solvable by a NDTM (with advice space $Z = \Cantor$) is Weak K\"onig's Lemma, i.e., the problem to find an infinite path through an infinite binary tree. The binary tree could be given via its characteristic function, so for any vertex we can decide whether or not its left and/or right child are present in the tree, too.

A NDTM solving Weak K\"onig's Lemma guesses a potential path through the tree (as an infinite binary sequence), and proceeds to output the guessed sequence, while simultaneously checking that every vertex used is actually present in the tree. If the path is invalid, it uses some non-existent vertex, which will be detected and result in the rejection of the guess. As every infinite binary tree has an infinite path, there is some valid path which will be given as output without rejection.

As long as the advice space is fixed to $Z = \Cantor$, any non-computable problem solvable by a corresponding NDTM is necessarily multi-valued. Using $Z = \Baire$ allows, for example, to write the Halting problem (in form of its characteristic sequence). One guesses, for each Turing machine-input pair, an upper bound on the runtime for a halting computation. If each number is correct, this allows to compute the Halting problem by simply simulating the computation for the given number of steps. On the other hand, if the input is incorrect, there will be a computation that halts after its allotted time has elapsed. This can be detected by continuing all simulations for ever.

Non-deterministic Type-2 machines are not proposed as a \emph{realistic} model of computation, but---just as in the classical case---as a useful conceptual tool helping to understand deterministic computations. A novel aspect is the availability of non-trivial theorems removing non-de\-ter-mi\-nism. Often, giving a non-de\-ter-mi\-nistic algorithm (with compact advice space) for a function is much easier than to directly come up with a deterministic algorithm. A metatheorem (\cite{brattka2}, \cite{paulybrattka}) then allows to remove the non-de\-ter-mi\-nism and to obtain a deterministic algorithm. Implicitly, such an approach is exhibited in \cite{rettinger} by \name{Rettinger} for computability of Jordan curves, and by \name{Galatolo}, \name{Hoyrup} and \name{Robas} in \cite{hoyrup} showing computability results for invariant measures. The present paper constitutes another application of this type.

\section{The representations}
In order to imbue \emph{a computable version of Theorem \ref{theo:jaynerogers}} with meaning, we need to clarify how the various objects are represented for the purpose of computations. The basic framework is computable analysis, laid out by \name{Weihrauch} in \cite{weihrauchd}. In particular, we will mostly work on computable metric spaces. In order to obtain a computable metric space from a separable metric space, one needs to fix a dense sequence, such that the distances are computable from the indices, this induces a computability structure on it. Every separable metric space is isomorphic to one admitting such a dense sequence, so in working on computable metric spaces we do not suffer any additional loss in generality as compared to working on separable metric spaces.

Representations of measurable sets and functions have been investigated by \name{Brattka} \cite{brattka}. The representations given below are straight-forward adoptions of those used by \name{Brattka}. As foundation for the representations, we obtain from represented spaces $\mathbf{X}$, $\mathbf{Y}$ the space $\mathcal{C}(\mathbf{X}, \mathbf{Y})$ of continuous functions from $\mathbf{X}$ to $\mathbf{Y}$, the product space $\mathbf{X} \times \mathbf{Y}$ and the spaces of closed $\mathcal{A}(\mathbf{X})$ and open $\mathcal{O}(\mathbf{X})$ subsets of $\mathbf{X}$. Regarding these constructions, see also \cite{pauly-synthetic-arxiv} by \name{Pauly} (based on \cite[Chapter 3.2]{paulyphd}).

\begin{definition}
Given a computable metric space $\mathbf{X}$, we define the space $\Delta_2^0(\mathbf{X})$ of $\Delta_2^0$ sets by identifying $(A_i, U_i)_{i \in \mathbb{N}} \in \mathcal{C}(\mathbb{N}, \mathcal{A}(\mathbf{X}) \times \mathcal{O}(\mathbf{X}))$ with the set $D := \bigcup_{i \in \mathbb{N}} A_i = \bigcap_{i \in \mathbb{N}} U_i$. If the two sets are unequal, the corresponding $(A_i, U_i)_{i \in \mathbb{N}}$ does not represent an element of $\Delta_2^0(\mathbf{X})$.
\end{definition}

\begin{definition}
\label{def:delta2}
Given a computable metric space $\mathbf{X}$ and a represented space $\mathbf{Y}$, we define the space $\Delta_2^0(\mathbf{X}, \mathbf{Y})$ of $\Delta_2^0$-measurable functions as the subspace of $\mathcal{C}(\mathcal{O}(\mathbf{Y}), \Delta_2^0(\mathbf{X}))$ containing functions of the form $f^{-1}$ for some $f : \mathbf{X} \to \mathbf{Y}$.
\end{definition}

This definition of the space $\Delta_2^0(\mathbf{X}, \mathbf{Y})$ adds a uniformity constraint: Not only do we require the preimage of any open set to be a $\Delta_2^0$-set, but we require the function mapping open sets to their preimages to be continuous itself. In the case of $\Sigma_n^0$-measurable functions discussed by \name{Brattka} in \cite{brattka}, this constraint actually comes for free: Whenever a function $g : \mathcal{O}(\mathbf{Y}) \to \Sigma_n^0(\mathbf{X})$ is of the form $g = f^{-1}$ for some function $f : \mathbf{X} \to \mathbf{Y}$, then $g$ is already continuous. For $\Delta_2^0$-measurable functions, the situation is more complicated, though (see Section \ref{section:classical}).

\begin{definition}
Given represented spaces $\mathbf{X}$, $\mathbf{Y}$ we define the space $\mathcal{C}^{\mathcal{A}-pw}(\mathbf{X}, \mathbf{Y})$ of $\mathcal{A}$-piecewise continuous functions by representing a function $f : \mathbf{X} \to \mathbf{Y}$ with a sequence\footnote{This occurrence of a dependent type can easily seen to be unproblematic.} $(A_i, f_i)_{i \in \mathbb{N}} \in \mathcal{C}(\mathbb{N}, \mathcal{A}(\mathbf{X}) \times \mathcal{C}(\mathbf{A}_i, \mathbf{Y}))$ such that $X = \bigcup_{i \in \mathbb{N}} A_i$ and $f_i = f_{|A_i}$.
\end{definition}

Unlike the situation for $\Delta_2^0(\mathbf{X}, \mathbf{Y})$, one can easily verify that for admissible $\mathbf{Y}$ the space $\mathcal{C}^{\mathcal{A}-pw}(\mathbf{X}, \mathbf{Y})$ contains extensionally exactly the functions classically consider piecewise continuous.

\section{Weihrauch reducibility and closed choice}
A convenient framework to discuss hypercomputability for general spaces is found in Weihrauch reducibility. Based on a related reducibility notion introduced by \name{Weihrauch} \cite{weihrauchb, weihrauchc}, it was primarily used by \name{Brattka}, \name{Gherardi}, \name{Marcone} and \name{Pauly} \cite{gherardi, paulyincomputabilitynashequilibria,brattka3,gherardi4} to pursue computable reverse mathematics. The resulting degree structure was investigated by \name{Brattka}, \name{Gherardi}, \name{Higuchi} and \name{Pauly} \cite{paulyreducibilitylattice,brattka2,paulykojiro}. Here we only reference the product operation $\times$, without making use of any specific properties.

\begin{definition}
For $f :\subseteq \mathbf{X} \mto \mathbf{Y}$, $g : \subseteq \mathbf{U} \mto \mathbf{Y}$, we say that $f$ is Weihrauch reducible to $g$ ($f \leq_\mathrm{W} g$), iff there are computable $H, K : \subseteq \Cantor \to \Cantor$, such that whenever $G : \subseteq \Cantor \to \Cantor$ is a realizer of $g$, we find $x \mapsto H(\langle x, GK(x)\rangle)$ to be a realizer of $f$.

We say that $f$ is strongly Weihrauch reducible to $g$ ($f \leq_{sW} g$), if there are computable $H, K : \subseteq \Cantor \to \Cantor$ such that whenever $G$ is a realizer of $g$, we find $HGK$ to be a realizer of $f$.
\end{definition}

Various important Weihrauch degrees are those of closed choice principles. For a represented space $\mathbf{X}$, we consider $C_\mathbf{X} : \subseteq \mathcal{A}(\mathbf{X}) \mto \mathbf{X}$ defined via $\dom(C_\mathbf{X}) = \mathcal{A}(\mathbf{X}) \setminus \{\emptyset\}$ and $x \in C_\mathbf{X}(A)$ iff $x \in A$. Closed choice principles are closely linked to non-de\-ter-mi\-nistic computation:

\begin{theorem}[\name{Brattka}, \name{de Brecht} \& \name{Pauly} {\cite[Theorem 7.2]{paulybrattka}}]
$f : \subseteq \mathbf{X} \mto \mathbf{Y}$ is non-de\-ter-mi\-nistically computable with advice space $\mathbf{Z}$ if and only if $f \leq_\mathrm{W} C_\mathbf{Z}$.
\end{theorem}

The class of multivalued functions reducible to $C_\mathbb{N}$ is not only also classified as those non-de\-ter-mi\-nistically computable with advice space $\mathbb{N}$, but also as those computable by a finitely revising machine (introduced by \name{Ziegler} \cite{ziegler2, ziegler3}) or by a generalized Turing machine allowed to make equality tests on $\Cantor$ (introduced by \name{Tavana} and \name{Weihrauch} \cite{tavana}) as can be seen following \cite{paulyoracletypetwo} by \name{Pauly}. In the present paper, we demonstrate that this class can be seen as a generalization of piecewise continuity to multivalued functions between represented spaces.

\section{The main result}
Our computable version of the Jayne-Rogers Theorem is based on the fact that evaluation for $\Delta_2^0$-measurable functions between computable metric spaces is non-de\-ter-mi\-nistically computable with advice space $\mathbb{N}$ (or alternatively with a finitely revising machine).
\begin{theorem}
\label{theo:eval}
Let $\mathbf{X}$, $\mathbf{Y}$ be computable metric spaces. The function $\textsc{Eval} : \Delta_2^0(\mathbf{X}, \mathbf{Y}) \times \mathbf{X} \to \mathbf{Y}$ satisfies $\textsc{Eval} \leq_\mathrm{W} C_{\mathbb{N}}$.
\begin{proof}
We show that $\textsc{Eval}$ is non-de\-ter-mi\-nistically computable with advice space $\mathbb{N} \times \Cantor$. By \cite[Theorem 7.2]{paulybrattka}, this implies $\textsc{Eval} \leq_\mathrm{W} C_{\mathbb{N} \times \Cantor}$. By \cite[Corollary 4.9]{paulybrattka} we have $C_{\mathbb{N} \times \Cantor} \leq_\mathrm{W} C_\mathbb{N} \times C_{\Cantor}$. As $\textsc{Eval}$ is single-valued, we can then invoke \cite[Theorem 5.1]{paulybrattka} to conclude $\textsc{Eval} \leq_\mathrm{W} C_\mathbb{N}$.

We regard $\mathbf{Y}$ as a subspace of the Hilbert cube $\mathbf{H} := \widehat{\uint}$, and assume the latter to be represented by the total representation $\delta_\mathbf{H} : \Cantor \to \widehat{\uint}$ (\cite[Proposition 4.1]{paulybrattka}), and the former by a suitable restriction $\delta_\mathbf{Y}$ of $\delta_\mathbf{H}$. The map $\operatorname{restrict} : \mathcal{O}(\mathbf{H}) \to \mathcal{O}(\mathbf{Y})$ defined via $\operatorname{restrict}(U) = U \cap Y$ is trivially computable.

A non-de\-ter-mi\-nistic algorithm to compute $\textsc{Eval}$ on input $f \in \Delta_2^0(\mathbf{X}, \mathbf{Y})$ and $x \in \mathbf{X}$ guesses $y \in \Cantor$ and $n \in \mathbb{N}$. The output in case of a successful guess is $y$.

As $\mathbf{H}$ is computably $T_2$, one can compute $\{\delta_\mathbf{H}(y)\} \in \mathcal{A}(\mathbf{H})$ from $y$, which has the same names as $H \setminus \{\delta_\mathbf{H}(y)\} \in \mathcal{O}(\mathbf{H})$. Hence, we can compute $Y \setminus \{\delta_\mathbf{Y}(y)\} = \operatorname{restrict}(H \setminus \{\delta_\mathbf{H}(y)\}) \in \mathcal{O}(\mathbf{Y})$, where we understand $\{\delta_\mathbf{Y}(y)\} = \emptyset$ for $y \notin \dom(\delta_\mathbf{Y})$.

 The information available on $f$ then allows to compute $f^{-1}(Y \setminus \{\delta_\mathbf{Y}(y)\}) \in \Delta_2^0(\mathbf{X})$, in particular, we can access $f^{-1}(Y \setminus \{\delta_\mathbf{Y}(y)\}) \in \Pi_2^0(\mathbf{X})$, and then also $f^{-1}(\{\delta_\mathbf{Y}(y)\}) \in \Sigma_2^0(\mathbf{X})$.

With $\mathbf{X}$ being a metric space, a $\sum_2^0$-set is the union of countably many closed sets, so we find $f^{-1}(\{\delta_\mathbf{Y}(y)\}) = \bigcup_{i \in \mathbb{N}} A_i$ with $A_i \in \mathcal{A}(\mathbf{X})$. We simultaneously test $x \in A_i$? for all $i \in \mathbb{N}$. If ever $x \notin A_i$ is confirmed for all $i \leq n$, the guess is rejected.

If a guess $(y, n)$ is never rejected, then there is some $i \leq n$ with $x \in A_i \subseteq  \left (\bigcup_{i \in \mathbb{N}} A_i \right ) = f^{-1}(\{\delta_\mathbf{y}\})$, hence $f(x) = \delta_\mathbf{Y}(y)$ and the output is correct. Also, if $\delta_\mathbf{Y}(y) = f(x)$, then $x \in \left (\bigcup_{i \in \mathbb{N}} A_i \right )$, so there is some $N \in \mathbb{N}$ with $x \in A_N$. But then $(y, N)$ can never be rejected.
\end{proof}
\end{theorem}

A multivalued function $f : \mathbf{X} \mto \mathbf{Y}$ reducible to $C_\mathbb{N}$ is clearly computable when restricted to those inputs where any fixed $n \in \mathbb{N}$ is a valid answer to the oracle question to $C_\mathbb{N}$. Any such set is a closed subset of $\dom(\delta_X)$, however, we will need the following lemma to lift these to closed subsets of $\mathbf{X}$.

Given a set $A \subseteq \Cantor$ and a represented space $\mathbf{X}$, we can obtain the represented space $\mathbf{X}_A$ by restricting the representation $\delta_X$ to $A$. This can be seen as a generalization of the subspace construction: If $A = \delta_X^{-1}\delta_X[A]$, then $\mathbf{X}_A = \mathbf{X}_{|\delta_X[A]}$ actually is a subspace of $\mathbf{X}$.

\begin{lemma}
\label{lemma:lifting}
Let $\mathbf{X}$, $\mathbf{Y}$ be computable metric spaces, and let $\delta_\mathbf{X}$ be proper. \begin{enumerate}
\item The map $\delta_X[ \ ] : \mathcal{A}(\Cantor) \to \mathcal{A}(\mathbf{X})$ is well-defined and computable.\footnote{Note that this does not follow directly from the fact that continuous images of compact sets are compact, together with compactness of $\Cantor$ and the $T_2$-property of $\mathbf{X}$ (cf.~\cite{pauly-synthetic-arxiv}), as $\delta_\mathbf{X}$ may very well be partial.}
\item Given $A \in \mathcal{A}(\Cantor)$ and $f \in \mathcal{C}(\mathbf{X}_A, \mathbf{Y})$ one can compute $f \in \mathcal{C}(\mathbf{X}_{|\delta_X[A]}, \mathbf{Y})$.
\end{enumerate}
\begin{proof}
\begin{enumerate}
\item As every computable metric space $\mathbf{X}$ is computably $T_2$, we can compute the compact singleton $\{x\}$ from $x \in \mathbf{X}$. As $\delta_X$ is assumed proper, $\delta_X^{-1}(\{x\})$ is compact, and moreover, can be computed from $x$ as a compact set. The intersection of a closed set and a compact set is uniformly compact, so from $x$ we compute $\delta_X^{-1}(\{x\}) \cap A$ as a compact set. Emptyness for compact sets is semidecidable, and $x \mapsto \operatorname{IsEmpty}(\delta_X^{-1}(\{x\}) \cap A)$ realizes $\delta_\mathbf{X}[A]$.
\item As before, we can compute $\delta_X^{-1}(\{x\}) \cap A$ from $x \in \mathbf{X}$, which just is the compact singleton $\{x\} \in \mathcal{K}(\mathbf{X}_{A})$. Images of compact sets under continuous functions are uniformly compact, so we obtain $\{f(x)\} \in \mathcal{K}(\mathbf{Y})$. Computable metric spaces are admissible, so from a compact singleton the value can be obtain, which gives us $f(x) \in \mathbf{Y}$. This treatment is uniform in $f$, thus yields the claim.
\end{enumerate}
\end{proof}
\end{lemma}

\begin{theorem}
\label{theo:piecewise}
Let $\mathbf{X}, \mathbf{Z}$ be computable metric spaces, and $\mathbf{Y}$ a represented space. If $f : \mathbf{X} \times \mathbf{Y} \to \mathbf{Z}$ satisfies $f \leq_\mathrm{W} C_\mathbb{N}$, then $y \mapsto (x \mapsto f(x, y)) : \mathbf{Y} \to \mathcal{C}^{\mathcal{A}-pw}(\mathbf{X}, \mathbf{Z})$ is computable.
\begin{proof}
Let the reduction $f \leq_\mathrm{W} C_\mathbb{N}$ be witnessed by $H$, $K$, and let $\mathcal{A}(\mathbb{N})$ be represented by $\psi^\mathbb{N}$. From $n \in \mathbb{N}$ and $q \in \dom(\delta_Y)$ one can compute $p \mapsto (n \in \psi^\mathbb{N}K(\langle p, q\rangle))$, which realizes a closed set $A_{n,q} \in \mathcal{A}(\Cantor)$. The map $p \mapsto \delta_ZH(\langle \langle p, q\rangle, n)$ now realizes $f_{A_{n,q}} \in \mathcal{C}(\mathbf{X}_{A_{n,q}}, \mathbf{Z})$. Using Lemma \ref{lemma:lifting}, both the closed cover and the partial realizers can be lifted from $\Cantor$ to $\mathbf{X}$.
\end{proof}
\end{theorem}

\begin{theorem}
\label{theo:cjrt}
For computable metric spaces $\mathbf{X}$, $\mathbf{Y}$ the maps $\id : \Delta_2^0(\mathbf{X}, \mathbf{Y}) \to \mathcal{C}^{\mathcal{A}-pw}(\mathbf{X}, \mathbf{Y})$ and $\id : \mathcal{C}^{\mathcal{A}-pw}(\mathbf{X}, \mathbf{Y}) \to  \Delta_2^0(\mathbf{X}, \mathbf{Y})$ are well-defined and computable.
\begin{proof}
The first part of the claim follows from combining Theorems \ref{theo:eval}, \ref{theo:piecewise}. For the second part, we need to show that from an open set $U \subseteq \mathbf{Y}$ and a piecewise continuous function $f : \mathbf{X} \to \mathbf{Y}$ we can compute the $\Delta_2^0$-set $f^{-1}(U)$.

Let $(A_i)_{i \in \mathbb{N}}$ be the closed cover of $\mathbf{X}$ coming with $f$. As any $f_{|A_i}$ is given as a continuous function, we can compute all $f^{-1}_{|A_i}(U) \in \mathcal{O}(\mathbf{A}_i)$. Invoking Lemma \ref{lemma:lifting} (1), we can extend the open set $f^{-1}_{|A_i}(U)$ to an open set $U_i \in \mathcal{O}(\mathbf{X})$, such that $f^{-1}_{|A_i}(U) = A_i \cap U_i$. In total, we find $f^{-1}(U) = \bigcup_{i \in \mathbb{N}} (A_i \cap U_i)$. In a computable metric space, any open set $U_i$ can be represented as a union $\bigcup_{j \in \mathbb{N}} \overline{U}_i^j$, hence we obtain $f^{-1}(U) = \bigcup_{i, j \in \mathbb{N}} A_i \cap \overline{U}_i^j$,

To express $f^{-1}(U)$ as a countable intersection of open sets, consider $f^{-1}(U^C)$. Any $f_{|A_i}^{-1}(U^C) =: B_i$ is closed in $A_i$, hence in $\mathbf{X}$. We have that $f^{-1}(U^C) = \bigcup_{i \in \mathbb{N}} B_i$ implies $f^{-1}(U) = \bigcap_{i \in \mathbb{N}} B_i^C = \bigcap_{i \in \mathbb{N}} (A_i^C \cup f_{|A_i}^{-1}(U))$, which is computable from the givens.
\end{proof}
\end{theorem}

\begin{corollary}
For computable metric spaces $\mathbf{X}$, $\mathbf{Y}$ we find that the map $\textsc{Eval} : \mathcal{C}^{\mathcal{A}-pw}(\mathbf{X}, \mathbf{Y}) \times \mathbf{X} \to \mathbf{Y}$ satisfies $\textsc{Eval} \leq_\mathrm{W} C_\mathbb{N}$.
\end{corollary}

We can call a function $f : \mathbf{X} \to \mathbf{Y}$ \emph{effectively} $\Delta_2^0$-measurable, if it has a \emph{computable} name in $\Delta_2^0(\mathbf{X}, \mathbf{Y})$. This means that one can compute $\Delta_2^0$-preimages of open sets without additional information. Likewise, a function with a computable name in $\mathcal{C}^{\mathcal{A}-pw}(\mathbf{X}, \mathbf{Y})$ is called \emph{piecewise computable}, this entails the existence of a countable cover by computably closed sets such that the corresponding restrictions are computable. As computable functions map computable elements to computable elements, we obtain the next corollary:

\begin{corollary}
\label{corr:effectivejaynerogers}
A function $f : \mathbf{X} \to \mathbf{Y}$ between computable metric spaces is effectively $\Delta_2^0$-measurable, if and only if it is piecewise computable, if and only if $f \leq_\mathrm{W} C_\mathbb{N}$.
\end{corollary}

\section{Markov-effective $\Delta_2^0$-measurability}
\label{section:Markov}
Besides the notion of effective $\Delta_2^0$-measurability used for Corollary \ref{corr:effectivejaynerogers}, there is a second possible definition. Given computable metric spaces, we can fix effective partial enumerations $(U_n)_{n \in \mathbb{N}}$ of the computably open subsets of $\mathbf{Y}$ and $(D_n)_{n \in \mathbb{N}}$ of the computably $\Delta_2^0$ sets in $\mathbf{X}$. Then we call $f : \mathbf{X} \to \mathbf{Y}$ \emph{Markov-effectively} $\Delta_2^0$-measurable, if there is a partial computable function $P : \subseteq \mathbb{N} \to \mathbb{N}$, such that whenever $U_n$ is an open subset of $\mathbf{X}$, we find $D_{P(n)}$ to be the $\Delta_2^0$ set $f^{-1}(U_n)$. This is the notion of effectivity that might be chosen in recursive analysis or effective descriptive set theory to capture $\Delta_2^0$-measurability.

We shall demonstrate that Markov-effective $\Delta_2^0$-measurability can be characterized via Weihrauch reducibility in a similar fashion to Corollary \ref{corr:effectivejaynerogers}, however, unlike its more uniform counterpart it is not related to piecewise computability (or finitely revising computability), but rather to \emph{low computability}.

Recall from recursion theory that $p \in \Cantor$ is called \emph{low}, iff the Turing jump of $p$ is the Halting problem, i.e., as simple as possible. In \cite{paulybrattka}, \name{Brattka}, \name{de Brecht} and \name{Pauly} suggested a uniform counterpart: We call $f : \mathbf{X} \mto \mathbf{Y}$ low computable, if from any name $p$ for $x \in \dom(f)$, we can compute a sequence converging to the Turing jump of some name $q$ for some $y \in f(x)$.

We restrict the considerations to Cantor space for now. Here, the effective enumeration $(U_n)_{n \in \mathbb{N}}$ of the computably open sets can be chosen total, and the Turing jump $J : \Cantor \to \Cantor$ defined via $J(p)(i) = 1$ iff $p \in U_i$.

\begin{theorem}
\label{theo:markov}
$f :\subseteq \Cantor \to \Cantor$ is Markov-effectively $\Delta_2^0$-measurable if and only if it is low computable.
\begin{proof}
\begin{description}
\item[$\Rightarrow$] Given a function $f$, if we have a uniform way of computing from an index $n$ a
$\Delta_2^0$-name for $f^{-1}(U_n)$, then we can create a limit computable function
that, on input $p$, converges to 1 if $f(p) \in U_n$ and converges to 0 if $f(p)
\not\in U_n$. So we can limit compute the $n$-th place of the Jump of
$f(p)$. If we can compute this uniformly in $n$, then by doing it all in parallel
we can low-compute $f$.
\item[$\Leftarrow$] On the other hand, if we can low-compute $f$, then we have a limit computable
function that converges to 1 if $f(p) \in U_n$ and 0 if $f(p) \not\in U_n$ (by just
looking at the $n$-th place of our low-computation of $f$). This implies that the
preimage of $U_n$ under $f$ is effectively $\Sigma_2^0$ (i.e., the effective union of countably many computably closed sets) and the preimage of the
complement of $U_n$ under $f$ is also effectively $\Sigma_2^0$, hence the preimage of $U_n$
is effectively $\Delta_2^0$.
\end{description}
\end{proof}
\end{theorem}

Low computability was characterized in \cite[Theorem 8.10]{paulybrattka} in terms of the function $\mathfrak{L} : \subseteq \mathcal{C}(\mathbb{N}, \Baire) \to \Baire$ defined via $\mathfrak{L}((p_i)_{i \in \mathbb{N}}) = q$ iff $\lim_{i \to \infty} p_i = J(q)$. One finds that any $f$ is low computable if and only if $f \leq_{sW} \mathfrak{L}$ holds.

\begin{corollary}
$f :\subseteq \Cantor \to \Cantor$ is Markov-effectively $\Delta_2^0$-measurable if and only if $f \leq_{sW} \mathfrak{L}$.
\end{corollary}

\begin{fact}
$\mathfrak{L} \nleq_\mathrm{W} C_\mathbb{N}$.
\end{fact}

\begin{corollary}
Markov-effective $\Delta_2^0$-measurability does not imply effective $\Delta_2^0$-measurability.
\end{corollary}

\begin{corollary}
There is a Markov-effectively $\Delta_2^0$-measurable function that is not even piecewise continuous.
\end{corollary}

We point out that \name{Higuchi} and \name{Kihara} \cite{kihara3} have independently obtained a similar result to Theorem \ref{theo:markov} which holds for Markov-effective $\Delta_n^0$-measurability and low$_n$-computability (although their terminology differs from ours).

\section{Computable vs.\ classical Jayne-Rogers theorem}
\label{section:classical}
As remarked after Definition \ref{def:delta2}, it is not guaranteed that the represented space $\Delta_2^0(\mathbf{X}, \mathbf{Y})$ actually contains {\bf all} $\Delta_2^0$-measurable functions. In principle, it is conceivable that some $f^{-1} : \mathcal{O}(\mathbf{Y}) \to \Delta_2^0(\mathbf{X})$ is well-defined, i.e., the inverse of a function $f : \mathbf{X} \to \mathbf{Y}$, yet lacks continuous realizers. As a consequence, the classical Jayne-Rogers Theorem does not follow directly from its computable counterpart.

However, for spaces $\mathbf{X}, \mathbf{Y}$ in its scope, the classical Jayne-Rogers Theorem states that all $\Delta_2^0$-measurable functions are elements in the space 
$\mathcal{C}^{\mathcal{A}-pw}(\mathbf{X}, \mathbf{Y})$---hence, by the computable Jayne-Rogers Theorem (Theorem \ref{theo:cjrt}), they are elements of $\Delta_2^0(\mathbf{X}, \mathbf{Y})$.

The classical Jayne-Rogers Theorem is not necessary for $\Delta_2^0(\mathbf{X}, \mathbf{Y})$ to contain all $\Delta_2^0$-measurable functions, though. Consider the space $\omega + 1 = \{A \in \mathcal{O}(\mathbb{N}) \mid n \in A \wedge m \geq n \Rightarrow m \in A\}$ with the Scott topology. Now $\mathcal{O}(\omega + 1)$ is countable, and there is a computable injection $\iota : \mathcal{O}(\omega + 1) \to \mathbb{N}^\triangledown$ ($\mathbb{N}^\triangledown$ are the natural number with the finitely revising representation). Using a list of all preimages of open sets, we find for any $\Delta_2^0$-measurable function $f : \Baire \to \omega + 1$ that $f^{-1} : \mathcal{O}(\omega + 1) \to (\Delta_2^0(\Baire))^\triangledown$ is continuous. As $(\Delta_2^0(\Baire))^\triangledown$ and $\Delta_2^0(\Baire)$ are isomorphic, $\Delta_2^0(\Baire, \omega + 1)$ encompasses all $\Delta_2^0$-measurable functions from $\Baire$ to $\omega + 1$.

Consider the function $e : \Baire \to (\omega + 1)$ defined via $e(p) = \{i \geq \max \{p(j) \mid j \in \mathbb{N}\} + 1\}$ if $\max \{p(j) \mid j \in \mathbb{N}\}$ exists and is even, $e(p) = \{i \geq \max \{p(j) \mid j \in \mathbb{N}\} - 1\}$ if $\max \{p(j) \mid j \in \mathbb{N}\}$ exists and is odd, and $e(p) = \mathbb{N}$ if $p$ is unbounded. Then $e$ is $\Delta_2^0$-measurable, yet not piecewise continuous. Thus, the Jayne-Rogers Theorem cannot be extended to $\omega + 1$ as codomain.

As just demonstrated, it is in principle possible to prove that the space $\Delta_2^0(\mathbf{X}, \mathbf{Y})$ does contain all $\Delta_2^0$-measurable functions for certain spaces $\mathbf{X}$, $\mathbf{Y}$ without resorting to the classical Jayne-Rogers Theorem. This gives hope that a more general result of this form, together with the computable Jayne-Rogers Theorem could be used as a simple proof for a (generalization of) the classical Jayne-Rogers Theorem.

It is worthwhile pointing out the analogy to the Kreitz-Weihrauch representation theorem for admissible representations proving that the space $\mathcal{C}(\mathbf{X}, \mathbf{Y})$ contains all the (topologically) continuous functions (see also \cite{schroder} by \name{Schr\"oder}, \cite{pauly-synthetic-arxiv} by \name{Pauly}). In the context of $\Sigma_n^0$-measurable functions, this connection has been explored in detail by \name{de Brecht} and \name{Yamamoto} in \cite{debrecht4}.

\section{Generalizing the main result}
The proof of Theorem \ref{theo:eval}, the center piece of our main result, makes no use of properties exclusive to metric spaces, and hence can be extended to more general spaces. The precise characterization of the suitable spaces is left for future work, however, we do have some limits how far the Jayne-Rogers Theorem can be extended.

In \cite{jaynerogers2}, \name{Jayne} and \name{Rogers} provide a counterexample with a metric, but non absolute Souslin-$\mathfrak{F}$ domain, and a discrete uncountable metric space as codomain assuming Martin's axiom. The latter is not available in a computable context, in particular, the absolute Souslin-$\mathfrak{F}$ condition is irrelevant for us.

A candidate condition is the $T_D$ separation axiom. A topological space is $T_D$, if any singleton is the intersection of an open and a closed set. A prototypic space failing the $T_D$ criterion is $\omega + 1$, hence the non-piecewise continuous function $e \in \Delta_2^0(\Baire, \omega + 1)$ introduced in the previous section bars an extension of the Jayne-Rogers Theorem to non-$T_D$ spaces.

On the other hand, a computable $T_D$ property suffices instead of the computable $T_2$ property employed in the proof of Theorem \ref{theo:eval}. A naive definition of computably $T_D$ requiring that from any singleton $x \in \mathbf{X}$ one can compute a pair $(A_x, U_x) \in \mathcal{A}(\mathbf{X}) \times \mathcal{O}(\mathbf{X})$ with $A_x \cap U_x = \{x\}$ turns out to be equivalent to computably $T_2$. However, allowing computation with finitely many mindchanges here, or, alternatively, requiring the computability of $x \mapsto \{x\} : \mathbf{X} \to \Delta_2^0(\mathbf{X})$ suffices for Theorem \ref{theo:eval}. The question which spaces embed into a computable $T_D$-space with a total Cantor-representation remains unresolved, though.

Another potential direction of generalization requires a better understanding of the interaction of two computational models, namely non-de\-ter\-mi\-nistic and limit machines. This could lead to a classification of the Weihrauch degree of function evaluation for functions where the preimages of open sets are $\Delta_n^0$ also for $n > 2$.

In a non-uniform way \name{Higuchi} and \name{Kihara} \cite{kihara3} made some progress in understanding the higher levels of effective measurability. They prove that a function where the preimages of $\Sigma_n^0$-sets effectively are $\Sigma_n^0$-sets will necessarily be non-uniformly computable.

In general, the results presented here could be an indication that \emph{computable descriptive set theory} could be developed relying heavily on Type-2 models of computation. That descriptive set theory has an underlying algorithmic structure is already evident from the r\^ole of games in this field, exhibited, e.g., in \cite{Wadge} by \name{Wadge} and \cite{semmes} by \name{Semmes}, which is generalized significantly in \cite{motto-ros3} by \name{Motto-Ros}. Computational models have some advantages over games, such as straight-forward closure properties under composition, that might provide additional usefulness to such an approach.

\bibliographystyle{eptcs}
\bibliography{pauly}
\end{document}